\documentclass[twocolumn,showpacs, showkeys,preprintnumbers,amsmath,amssymb]{revtex4}

\usepackage{graphicx}
\usepackage{dcolumn}
\usepackage{bm}
\usepackage{enumerate}

\usepackage{rotating}

\begin{document}


\title{The trap DOS in small molecule organic semiconductors: \\A quantitative comparison of thin-film transistors with single crystals}

\author{Wolfgang L. Kalb}%
 \email{kalb@phys.ethz.ch}
\author{Simon Haas}
\author{Cornelius Krellner}
\author{Thomas Mathis}
\author{Bertram Batlogg}
\affiliation{%
Laboratory for Solid State Physics, ETH Zurich, 8093 Zurich,
Switzerland
}%

\date{\today}

\begin{abstract}
We show that it is possible to reach one of the ultimate goals of
organic electronics: producing organic field-effect transistors with
trap densities as low as in the bulk of single crystals. We studied
the spectral density of localized states in the band gap (trap DOS)
of small molecule organic semiconductors as derived from electrical
characteristics of organic field-effect transistors or from
space-charge-limited-current measurements. This was done by
comparing data from a large number of samples including thin-film
transistors (TFT's), single crystal field-effect transistors
(SC-FET's) and bulk samples. The compilation of all data strongly
suggests that structural defects associated with grain boundaries
are the main cause of ``fast'' hole traps in TFT's made with
vacuum-evaporated pentacene. For high-performance transistors made
with small molecule semiconductors such as rubrene it is essential
to reduce the dipolar disorder caused by water adsorbed on the gate
dielectric surface. In samples with very low trap densities, we
sometimes observe a steep increase of the trap DOS very close
($<0.15$\,eV) to the mobility edge with a characteristic slope of
$10-20$\,meV. It is discussed to what degree band broadening due to
the thermal fluctuation of the intermolecular transfer integral is
reflected in this steep increase of the trap DOS. Moreover, we show
that the trap DOS in TFT's with small molecule semiconductors is
very similar to the trap DOS in hydrogenated amorphous silicon even
though polycrystalline films of small molecules with van der
Waals-type interaction on the one hand are compared with covalently
bound amorphous silicon on the other hand. Although important
conclusions can already be drawn from the existing data, more
experiments are needed to complete the understanding of the trap DOS
near the band edge in small molecule organic semiconductors.
\end{abstract}

\pacs{73.20.Hb, 73.61.Ph, 73.20.At, 85.30.De}
\keywords{organic semiconductor, pentacene, rubrene, trap density of states, organic field-effect transistor}
\maketitle

\section{Introduction}

Organic semiconductors are envisioned to revolutionize display and
lighting technology. The remaining engineering-related challenges
are being tackled and the first products are commercially available
already. To guarantee a sustainable market entry, however, it is
important to further deepen the understanding of organic
semiconductors and organic semiconductor devices. Electronic trap
states in organic semiconductors severely affect the performance of
such devices. For organic thin-film transistors (TFT's), for
example, key device parameters such as the effective charge
mobility, the threshold voltage, the subthreshold swing as well as
the electrical and environmental stability are severely affected by
trap states at the interface between the gate dielectric and the
semiconductor. Trap states in organic semiconductors have been
studied for several decades.\cite{PopeM} Although the first organic
field-effect transistors emerged in the 1980's, (polymeric
semiconductors: Ref.~\onlinecite{EbisawaF1983}, small molecule
organic semiconductors: Ref.~\onlinecite{HorowitzG1989}) it is only
recently, that trap states in organic field-effect transistors are a
subject of intense scientific investigation (Refs.~\cite{BragaD2009,
SirringhausH2009, KalbWL2010} and references therein).

The present study is focused on trap densities in small molecule
organic semiconductors. These solids consist of molecules with
loosely bound $\pi$-electrons. The $\pi$-electrons are transferred
from molecule to molecule and, therefore, are the source of charge
conduction. Small molecule organic semiconductors tend to be
crystalline and can be obtained in high purity. Typical materials
are oligomers such as pentacene, tetracene or sexithiophene but this
class of materials also includes e.g. rubrene, C$_{60}$ or the
soluble material TIPS pentacene (Ref.~\onlinecite{ParkSK2007}).

Trap densities are often given as a volume density thus averaging
over various trapping depths. The spectral density of localized
states in the band gap, i.e. the trap densities as a function of
energy (trap DOS), gives a much deeper insight into the charge
transport and device performance. In this paper we compare, for the
first time, the trap DOS in various samples of small molecule
organic semiconductors including thin-film transistors (TFT's) where
the active layer generally is polycrstalline and organic single
crystal field-effect transistors (SC-FET's). These data are also
compared with the trap DOS in the bulk of single crystals made of
small molecule semiconductors. It turns out that it is this
comparison of trap densities in TFT's, SC-FET's and in the bulk of
single crystals that is particularly rewarding.

The trap DOS in organic semiconductors can be derived from several
different experimental techniques, including measurements of
field-effect transistors, space-charge-limited current (SCLC)
measurements, thermally stimulated currents (TSC), Kelvin-probe,
time-of-flight (TOF) or capacitance measurements. For the present
study, we focus on the trap DOS as derived from electrical
characteristics of organic field-effect transistors or from SCLC
measurements of single crystals.

We begin with a brief discussion of charge transport in small
molecule semiconductors followed by a summary of the current view of
the origin of trap states in these materials. After a comparison of
different methods to calculate the trap DOS from electrical
characteristics of organic field-effect transistors we are
eventually in a position to compile, compare and discuss trap DOS
data.

\section{\label{section-chargetransport}Charge transport in small molecule organic semiconductors}

Even in ultrapure single crystals made of small molecule
semiconductors, the charge transport mechanism is still
controversial. The measured mobility in ultrapure crystals increases
as the temperature is decreased according to a power law
$\mu_{0}\propto T^{n}$.\cite{WartaW1985} This trend alone would be
consistent with band transport. However, the mobilities $\mu_{0}$ at
room temperature are only around 1\,cm$^{2}$/Vs and the estimated
mean free path thus is comparable to the lattice constants. It has
often been noticed that this is inconsistent with band
transport.\cite{WartaW1985}

Since the molecules in the crystal have highly polarizable
$\pi$-orbitals, polarization effects are not negligible in a
suitable description of charge transport in organic semiconductors.
\textit{Holstein's} polaron band model considers electron-electron
interactions and the model has recently been
extended.\cite{HolsteinT1959Part1, HolsteinT1959Part2,
HannewaldK2004} With increasing temperature, the polaron mass
increases. This effect is accompanied by a bandwidth narrowing and
inevitably results in a localization of the charge carrier.
Consequently, this model predicts a transition from band transport
at low temperature to phonon-assisted hopping transport at higher
temperatures (e.g. room temperature). The model may explain the
experimentally observed increase in mobility with decreasing
temperature and seems to be consistent with the magnitude of the
measured mobilities at room temperature.

On the other hand, thermal motion of the weakly bound molecules in
the solid is large compared to inorganic crystals. Such thermal
motions most likely affect the intermolecular transfer integral.
Indeed, \textit{Troisi et al.} have shown that, at least for
temperatures above 100\,K, the fluctuation of the intermolecular
transfer integral is of the same order of magnitude as the transfer
integral itself in materials such as pentacene, anthracene or
rubrene.\cite{TroisiA2006a, TroisiA2007} As a consequence, the
fluctuations do not only introduce a small correction, but determine
the transport mechanism and limit the charge carrier
mobility.\cite{TroisiA2006b} Clearly, the thermal fluctuations are
less severe at a reduced temperature and the calculations predict a
mobility that increases with decreasing temperature, according to a
power law. This is in excellent agreement with the measured
temperature-dependence in ultrapure crystals. Moreover, the model
predicts mobilities at room temperature between 0.1\,cm$^{2}$/Vs and
50\,cm$^{2}$/Vs, which also is in good agreement with
experiment.\cite{TroisiA2006c, TroisiA2007} Interestingly, the
importance of thermal disorder is supported by recent tetrahertz
transient conductivity measurements on pentacene
crystals.\cite{LaarhovenHA2008}

\begin{figure}
\includegraphics[width=0.70\linewidth]{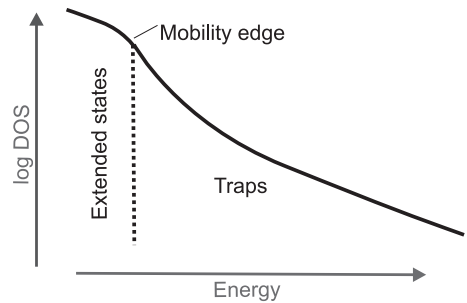}
\caption{\label{figure-DOSsketch} Schematic representation of the
mobility edge separating localized states (traps) from extended
states. At the mobility edge, the mobility as a function of energy
abruptly rises and only the charge carriers that are thermally
activated to states above the mobility edge contribute to the charge
transport.}
\end{figure}

In essence, the band broadening due to the thermal motion of the
molecules is expected to result in electronic trap states which
would be related to the intrinsic nature of small molecule
semiconductors.\cite{SleighJP2009} Clearly, trap states can also be
due to extrinsic defects and these traps can completely dominate the
charge transport resulting in an effective mobility
$\mu_{eff}$.\cite{HorowitzG1995, SchauerF1999} For amorphous
inorganic semiconductors such as amorphous silicon, the mobility
edge picture has been developed
(Fig.~\ref{figure-DOSsketch}).\cite{AndersonPW1972, MarshallJM1983}
The mobility edge separates extended from localized states. The
existence of extended states in amorphous silicon is attributed to
the similarity of the short-range configuration of the atoms in the
amorphous phase which is similar to the configuration in the
crystalline phase.\cite{MarshallJM1983} Hopping in localized states
is expected to be negligible if transport in extended states exists,
i.e. we have an abrupt increase in mobility at the mobility edge.
Only the charge carriers that are thermally activated to states
above the mobility edge contribute to the transport of charge.

In the following we assume that charge transport in small molecule
semiconductors can be described by an effective transport level and
a distribution of trap states below this transport level. The
mobility edge model is a specific realization of this very general
assumption. In a \textit{completely} disordered material (no
short-range order) all electronic states are
localized.\cite{MarshallJM1983} The charge carriers are highly
localized and hop from one molecule to the next. However, even this
situation can be described by introducing an effective transport
level and a broad distribution of trap states below the transport
level.\cite{ArkhipovVI2003} In the following, we use the term
valence band edge. This term may generally be interpreted as the
effective transport level and denotes the mobility edge in the
mobility edge picture.

\section{\label{section-origin}Causes of trap states in small molecule organic semiconductors}

We proceed by summarizing the current view of the microscopic origin
of traps in small molecule semiconductors. Charge carrier traps
within the semiconductor are caused by structural defects or
chemical impurities. Chemical impurities may also cause a
surrounding of structural defects by distorting the host
lattice.\cite{ProbstKH1975} On the other hand, chemical impurities
tend to accumulate in regions with increased structural disorder
(Ref.~\onlinecite{PopeM}) as well as at the surface of a crystal
(Ref.~\onlinecite{JurchescuOD2007}). Trap states caused by the gate
dielectric can become very important in organic field-effect
transistors. Finally, as mentioned already, also the thermal
fluctuations of the molecules are expected to result in shallow trap
states within the band gap.

\subsection{Structural defects}

In the bulk of ultrapure anthracene or naphthalene crystals, typical
densities of vacancies (a dominant point defect) are of the order of
$10^{14}-10^{15}$\,cm$^{-3}$ (Ref.~\cite{PopeM}, p.~222). Vacancies
are expected to be concentrated close to other structural defects
due to a reduced formation energy.\cite{PopeM} Extended structural
defects (e.g. edge dislocations, screw dislocations or low-angle
grain boundaries) can be present in significant densities in organic
crystals, e.g. 10$^{19}$\,cm$^{-3}$ (Ref.~\cite{PopeM}, p.226).
Therefore, extended structural defects are thought to be the main
source of traps in ultrapure organic crystals.\cite{OwenGP1974}

Thin films of small molecule semiconductors are expected to have a
higher density of structural defects than single crystals.
Thin films of small molecule semiconductors are often
polycrystalline and grain boundaries can limit the charge transport
in such films. For example, measurements of sexithiophene-based
transistors with SiO$_{2}$ gate dielectric and an active channel
consisting of only two grains and one grain boundary show, that the
transport is in fact limited by the grain
boundary.\cite{ChwangAB2001} At the grain boundary, a high density
of traps exists and the density of these traps per unit area of the
active accumulation layer is of the order of
$10^{12}$\,cm$^{-2}$.\cite{ChwangAB2001}

In the following, we focus on structural defects in vacuum
evaporated pentacene films which are of particular relevance for
this work. Since pentacene films are often polycrystalline, large
angle grain boundaries are expected to produce additional structural
defects also in this material. The effect of grain boundaries on
charge transport in pentacene films is still controversial. Atomic
force measurements (AFM) of ultrathin pentacene films have clearly
shown, that the field-effect mobility in pentacene-based transistors
can be higher in films with smaller grains.\cite{KalbW2004} In
addition, some experimental evidence indicates that there is no
correlation between charge trapping and topographical features in
pentacene thin films.\cite{MullerEM2005} On the contrary, it has
recently been shown that long-lived (energetically deep) traps that
cause gate bias stress effects in pentacene-based TFT's are mainly
concentrated at grain boundaries.\cite{TelloM2008} Another important
cause of structural disorder in pentacene films is polymorphism
since pentacene can crystallize in at least four different
structures (phases). It is quite common that at least two of these
phases coexist in pentacene thin films.\cite{MatheusCC2001,
DimitrakopoulosCD2002, ChengHL2007, SiegristT2007}

A theoretically study deals with in-grain defects in vacuum
evaporated pentacene films.\cite{VerlaakS2007} Structural defects
are formed during the film growth. Upon addition of more and more
``defective'' molecules at a given site, the ideal crystal structure
becomes energetically more and more favourable. The system
eventually relaxes into the ideal crystal structure during the
continuation of the film growth. The relaxation happens, provided
that the evaporation rate is low enough and that there is enough
time for relaxation.\cite{VerlaakS2007} In this study it is
suggested that structural defects within the grains of a pentacene
film that resist relaxation cannot exceed densities of
10$^{16}$\,cm$^{-3}$, at typical growth conditions. A structural
defect can, however, influence the electronic levels of 10
surrounding molecules even if these molecules are in the perfect
crystal configuration. It is concluded that grain boundaries (and
not in-grain defects) are the most prominent cause of structural
defects in pentacene thin films.\cite{VerlaakS2007}

On the other hand, an experimental study identifies pentacene
molecules that are displaced slightly out of the molecular layers
that make up the crystals.\cite{KangJH2005} By means of high
impedance scanning tunneling microscopy (STM), specific defect
islands were detected in pentacene films with monolayer coverage.
Within the defect islands, the pentacene molecules are displaced up
to 2.5\,{\AA} along the long molecular axis out of the pentacene
layer with a broad distribution in the magnitude of the
displacements. Electronic structural calculations show that the
displaced molecules lead to traps for both electrons and holes. The
maximum displacement of the pentacene molecules as seen by STM is
2.5\,{\AA} and this corresponds to a maximum trap depth of
0.1\,eV.\cite{KangJH2005}

\subsection{Chemical impurities}

The best method to produce crystals of small molecule semiconductors
includes a zone refinement step in the purification procedure
(Ref.~\cite{PopeM}, p.~224). Even such crystals still have a
considerable impurity content. Anthracene, for example, still has an
impurity content of 0.1\,ppm in the best crystals, which corresponds
to a volume density of $\approx10^{14}$\,cm$^{-3}$
(Ref.~\cite{PopeM}, p.~224). Zone refinement produces organic
materials of much higher purity as compared to purification by
sublimation.\cite{PflaumJ2006} However, zone refinement can only be
applied if the material can be molten without a chemical reaction or
a decomposition to occur. This is not possible for many materials
including tetracene or pentacene. Thus, much higher impurity
concentrations are expected e.g. in tetracene or
pentacene.\cite{PflaumJ2006} An experimental study indicates that in
tetracene single crystals the charge carrier mobility is limited by
chemical impurities rather than by structural
defects.\cite{PflaumJ2006}

The ability of a chemical impurity to act as trap depends on its
accessible energy levels. In a simplistic view a hole trap forms if
the ionization energy of the impurity is smaller than the ionization
energy of the host material.\cite{PopeM} We focus on pentacene, and
the center ring of the pentacene molecule is expected to be the most
reactive.\cite{MattheusCC2002, NorthrupJE2003, KnippD2009} An
important impurity is thus thought to be the oxidized pentacene
species 6,13-pentacenequinone, where two oxygen atoms form double
bonds with the carbon atoms at the 6,13-positions. According to
theoretical studies, Pentacenequinone is expected to lead to states
in the band gap of pentacene (Ref.~\onlinecite{NorthrupJE2003} and
\onlinecite{KnippD2009}) and may predominantly act as scattering
center (Ref.~\onlinecite{JurchescuOD2004}). Repeated purification of
pentacene by sublimation can result in very high mobilities in
pentacene single crystals.\cite{JurchescuOD2004} Another common
impurity in pentacene is thought to be 6,13-dihydropentacene, where
additional hydrogen atoms are bound both at the 6- and at the
13-position.\cite{MattheusCC2002}

\subsection{Trap states due to the gate dielectric}

Properties of the gate dielectric's surface such as surface
roughness, surface free energy and the presence of heterogenous
nucleation sites are expected to play a key role in the growth of
small molecule semiconductor films from the vapour phase thus
influencing the quality of the films. Apart from growth-related
effects, the sole presence of the gate dielectric can influence the
charge transport in a field-effect transistor especially because the
charge is transported in the first few molecular layers within the
semiconductor at the interface between the gate dielectric and the
semiconductor. Thus, also FET's based on single crystals are
affected, even laminated (flip-crystal-type) SC-FET's.

\subsubsection{Chemical nature of the gate dielectric}

The surface of the gate dielectric contains chemical groups that act
as charge carrier traps. The trapping mechanism may be as simple as
the one discussed above for chemical impurities. This means that the
trapping depends on the specific surface chemistry of the gate
dielectric but the ability of certain chemical groups on the surface
of the gate dielectric to cause traps will also depend on the nature
of the small molecule semiconductor. The trapping mechanism can also
be seen as a reversible or irreversible electrochemical reaction
driven by the application of a gate voltage.\cite{NorthrupJE2003}
Chemical groups on the surface of the gate dielectric certainly
affect the transport of electrons in n-type field-effect
transistors.\cite{AhlesM2004, ChuaLL2005, YoonMH2006}

\subsubsection{Adsorbed water}

Water adsorbed on the gate dielectric may dissociate and react with
pentacene. One possible reaction product is 6,13-dihydropentacene.
The number of impurities that are formed can depend on the
electrochemical potential and would thus increase as the gate
voltage is ramped up in a field-effect
transistor.\cite{NorthrupJE2003}

It has also been suggested that water causes traps by reacting with
the surface of the gate dielectric. Water on a SiO$_{2}$ gate
dielectric with a large number of silanol groups (-Si-OH) causes the
formation of SiO$^{-}$-groups and the latter groups can act as hole
traps.\cite{KumakiD2008}

In addition to chemical reactions involving water, water molecules
may act as traps themselves just like any other chemical impurity. A
polar impurity molecule leads to an electric field dependent trap
depth though.\cite{SworakowskiJ1999}

Even if a polar impurity does not lead to a positive trap depth, its
dipole moment modifies the local value of the polarization energy
since we have highly polarizable $\pi$-orbitals in organic
semiconductors. This results in traps in the vicinity of the water
molecules.\cite{SworakowskiJ1999, VerlaakS20072} The net effect is a
significant broadening of the trap DOS at the
insulator-semiconductor interface.\cite{SworakowskiJ1999}

\subsubsection{Dielectric constant of the gate dielectric}

It has been suggested that the polarity of the gate dielectric
surface impedes the charge transport as described in the
following.\cite{VeresJ2003, VeresJ2004} A more polar surface has
randomly oriented dipoles which lead to a modification of the local
polarization energy within the semiconductor and thus to a change of
the site energies. As in the case of polar water molecules, this
brings a broadening of the trap DOS. The dependence of the mobility
on the dielectric constant of the gate dielectric has been observed
with conjugated polymers (Refs.~\onlinecite{VeresJ2003} and
\onlinecite{VeresJ2004}) and with rubrene single crystal
field-effect transistors.\cite{StassenAF2004} More recently, a model
has been put forward to quantitatively study the effect of randomly
oriented static dipole moments within the gate
dielectric.\cite{RichardsT2008} The model predicts a significant
broadening of the trap DOS within the first 1\,nm at the
insulator-semiconductor interface and can explain the dependence of
the mobility on the dielectric constant of the gate dielectric
quantitatively.\cite{RichardsT2008} In this context, it is important
to realize that surfaces with a low polarity have a low surface free
energy and are thus expected to have a high water repellency as
well. Clearly, the high water repellency leads to a a reduced amount
of water at the critical insulator-semiconductor
interface.\cite{VeresJ2004}

\subsection{Thermal motion of the molecules}

As already mentioned in Sec.~\ref{section-chargetransport}, the
thermal fluctuations of the intermolecular transfer integral may be
of the same order of magnitude as the transfer integral itself in
small molecule semiconductors such as pentacene, anthracene or
rubrene.\cite{TroisiA2006a, TroisiA2007} A theoretical study has
pointed out that the large fluctuations in the transfer integral
result in a tail of trap states extending from the valence band edge
into the gap.\cite{SleighJP2009} Moreover, the band tail is
temperature-dependent. The extension of the band tail increases with
temperature due to an increase in the thermal motion of the
molecules.\cite{SleighJP2009} For pentacene the theoretical study
predicts exponential band tails $N=N_{0}\exp(-E/E_{0})$ with
$E_{0}=12.7$\,meV at $T=300$\,K and $E_{0}=6.9$\,meV at $T=100$\,K.
Some experimental evidence suggests, that trap states due to the
thermal motion of the molecules play a role in samples with a low
trap density.\cite{KrellnerC2007, HaasS20072, PernstichKP2008}

\section{\label{section-quanti}Calculating the trap DOS from experiment}

\begin{figure}
\includegraphics[width=0.90\linewidth]{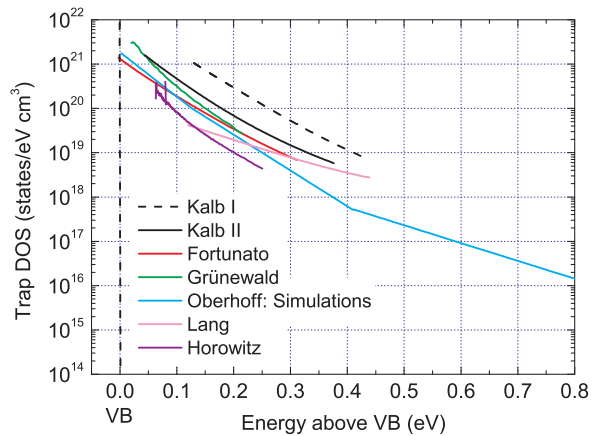}
\caption{\label{figure-compmethods} (Color online) Spectral density
of localized states in the band gap (trap DOS) of pentacene as
calculated with several methods from the same set of transistor
characteristics. The transistor characteristics were measured with a
pentacene-based TFT employing a polycrystalline pentacene film and a
SiO$_{2}$ gate dielectric. The energy is relative to the valence
band edge (VB). The choice of the method to calculate the trap DOS
has a considerable effect on the final result. Adapted from
Ref.~\onlinecite{KalbWL2010}.}
\end{figure}

Field-effect transistors are often used to measure the trap DOS. The
trap DOS can be calculated from the measured transfer
characteristics with various analytical methods or by simulating the
transistor characteristics with a suitable computer program. In
Sec.~\ref{section-comparison} we quantitatively compare the trap DOS
from various studies in the literature with our data. Since in these
studies different methods were used to derive the trap DOS, it is
necessary to ensure that all these methods lead to comparable
results. Analytical methods that are relevant for the comparison in
Sec.~\ref{section-comparison} were developed by \textit{Lang et al.}
(Ref.~\onlinecite{LangDV2004}), \textit{Horowitz et al.}
(Ref.~\onlinecite{HorowitzG1995}), \textit{Fortunato et al.}
(Ref.~\onlinecite{FortunatoG1988}), \textit{Gr\"unewald et al.}
(Ref.~\onlinecite{GrunewaldM1980}) and \textit{Kalb et al.} (method
I: Ref.~\onlinecite{KalbWL2008}, method II:
Ref.~\onlinecite{KalbWL2010}). The trap DOS as calculated with the
different methods from the same set of measured data is shown in
Fig.~\ref{figure-compmethods}.\cite{KalbWL2010} Clearly, the choice
of the method to calculate the trap DOS has a considerable effect on
the final result. The graph also contains the trap DOS obtained by
simulating the transistor characteristics with a computer program
developed by \textit{Oberhoff et al.} and this may be seen as the
most accurate trap DOS.\cite{OberhoffD2007, KalbWL2010} The
analytical results agree to a varying degree with the simulation.
Method I by \textit{Kalb et al.} gives a good estimate of the slope
of the trap DOS but overestimates the magnitude of the trap
densities which can be attributed to a neglect of the
temperature-dependence of the band mobility
$\mu_{0}$.\cite{KalbWL2010} For the method by \textit{Lang et al.},
the effective accumulation layer thickness $a$ is assumed to be
constant (gate-voltage independent). An effective accumulation layer
thickness of $a=7.5$\,nm is generally used. The method by
\textit{Lang et al.} leads to a significant underestimation of the
slope of the trap DOS and, with an effective accumulation layer
thickness of $a=7.5$\,nm, to a significant underestimation of the
trap densities very close to the valence band edge (VB). These
deviations need to be considered in the following analysis.

\section{\label{section-comparison}Comparison of trap DOS data}

On the one hand, trap DOS data were taken from publications by
various groups that are active in the field. The data were extracted
by using the Dagra software which allows to convert plotted data
e.g. in the figures of PDF files into data columns. On the other
hand, we also add to the following compilation unpublished data from
experiments in our laboratory.

We focus on the trap DOS in small molecule semiconductors. Since
almost no data exists in the literature on the trap DOS in
solution-processed small molecule semiconductors, we almost
exclusively deal with the trap DOS in vapour-deposited small
molecules. More specifically, the data are from TFT's which were
made by evaporating the small molecule semiconductors in high
vacuum. The single crystals for the SC-FET's and for the
measurements of the bulk trap DOS were grown by physical vapour
transport (sublimation and recrystallization in a stream of an inert
carrier gas).\cite{LaudiseRA1998} Moreover, the electron trap DOS
close to the conduction band edge (CB) has rarely been studied so
far in small molecule semiconductors and, with one exception, we are
dealing with the hole trap DOS in small molecule semiconductors in
the following.

\subsection{Trap DOS from TFT's}

\begin{figure}
\includegraphics[width=0.90\linewidth]{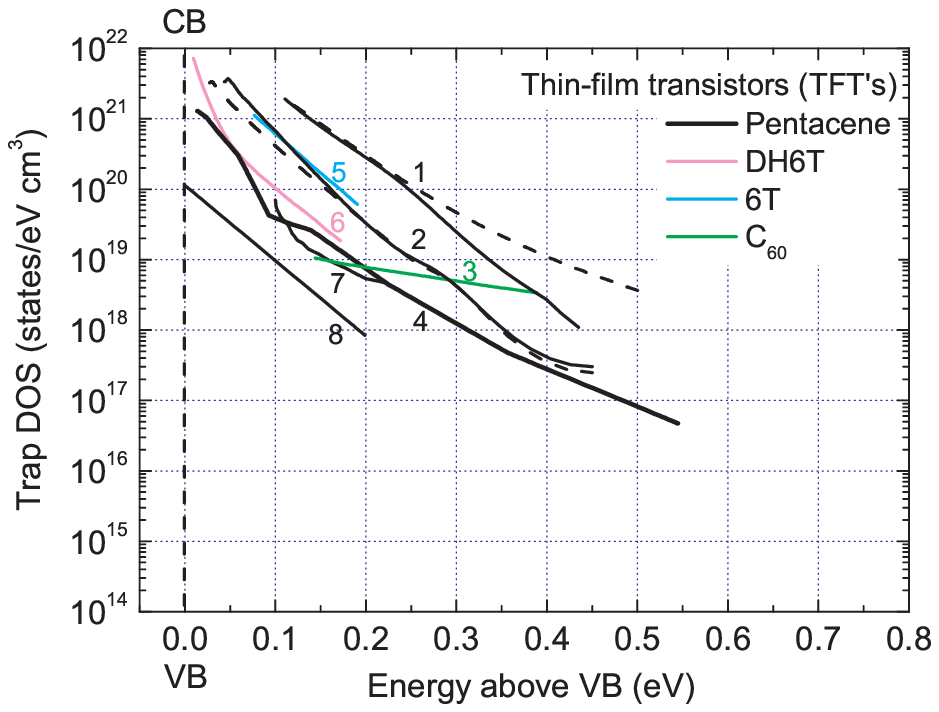}
\caption{\label{figure-TFTs} (Color online) Trap DOS from thin-film
transistors (TFT's) made with small molecule organic semiconductors.
Several different semiconductors, gate dielectrics and methods to
calculate the trap DOS were used. Some details of the TFT
fabrication are listed in Table~\ref{table-TFTs} along with the
method that was used to calculate the trap DOS and the reference of
the data. Small molecule semiconductors tend to be crystalline and
can be obtained in high purity. Typical materials are oligomers such
as pentacene or sexithiophene but this class of materials also
includes e.g. rubrene or C$_{60}$. The molecules interact by weak
van der Waals-type forces and have loosely bound $\pi$-electrons
which are the source of charge conduction.}
\end{figure}

\begin{table*}
\caption{\label{table-TFTs} Thin-film transistors (TFT's) made with
small molecule organic semiconductors: details and references of the
data shown in Fig.~\ref{figure-TFTs}. The purity of the small
molecule semiconductor (starting material), the nature of the gate
dielectric as well as contact effects may influence the trap DOS.
The choice of the method to calculate the trap DOS has a
considerable effect on the final result and is also listed.}
\begin{ruledtabular}
\begin{tabular}{llllllllll}
Data no. & Semiconductor & Starting material & Gate dielectric &
Contact & Contact & Method &
Comment & Ref. data\\
&  &  & & material & type & &&&  \\  \hline

1 & Pentacene & Aldrich (purum), & SiO$_{2}$ & Au &
TC\footnotemark[1]\footnotemark[2] & \textit{Kalb I} & Full line: &
\cite{KalbWL2008}\\
& & $2\times$ recrystallized  & & & & & pristine sample,
& \\
& & & & & & &dashed line:
& \\
& & & & & & &after O$_{2}$ exposure
& \\

2 & Pentacene & Aldrich (purum), & SiO$_{2}$ & Au &
TC\footnotemark[2] & \textit{Gr\"unewald }& Full line: &
\cite{KalbWL2007}\\
& & $2\times$ recrystallized & & & & & pristine sample,
& \\
& & & & & & &dashed line:
& \\
& & & & & & &after aging
& \\

3 & C$_{60}$ &  & SiO$_{2}$ & Au & TC & \textit{Lang} & Electron
traps, &
\cite{KawasakiN2007}\\
& & & & & & &$a=7.5$\,nm\footnotemark[3] & \\

4 & Pentacene & Aldrich (97\%), & PMMA\footnotemark[4] & Au & TC &
\textit{Fortunato} &  &
\cite{DeAngelisF2006, DeAngelisF2005}\\
& & no additional & buffer layer & &&&\\
& & purification & on SiO$_{2}$  & &&&\\

5 & 6T\footnotemark[5] &  & PMMA & Au & TC & \textit{Horowitz }& &
\cite{HorowitzG1995}\\

6 & DH6T\footnotemark[6] &  & PMMA & Au & TC & \textit{Horowitz} & &
\cite{HorowitzG1995}\\

7 & Pentacene & & SiO$_{2}$ & Au & TC & \textit{Lang }& Pristine
sample, &
\cite{VanoniC2009}\\
& & & & & & &$a=10$\,nm & \\

8 & Pentacene & Aldrich (97\%), & SiO$_{2}$ & Au & TC &
\textit{V\"olkel}\footnotemark[7] &
 & \cite{VoelkelAR2002, KnippD2001} \\
& & no additional & & & & &  & \\
& & purification & & & & & & \\
\end{tabular}
\end{ruledtabular}
\footnotetext[1]{TC = top contacts.} \footnotetext[2]{Gated
four-terminal measurements.} \footnotetext[3]{$a$ is the constant
effective accumulation layer thickness used for the calculations.}
\footnotetext[4]{Polymethylmetacrylate.}\footnotetext[5]{Sexithiophene.}\footnotetext[6]{Substituted
dihexyl-sexithiophene.} \footnotetext[7]{Computer simulations.}

\end{table*}

In Fig.~\ref{figure-TFTs} we show the trap DOS in various TFT's made
with small molecule semiconductors. All transistors were fabricated
by evaporating the organic material onto substrates comprising the
gate electrode and gate dielectric and were completed by evaporating
Au top contacts (TC). Details of the data are given in
Table~\ref{table-TFTs}. Apart from one exception we are dealing with
hole traps that are plotted relative to the valence band edge (VB).
The exception is $C_{60}$ (green line in Fig.~\ref{figure-TFTs}) and
the electron trap densities are plotted relative to the conduction
band edge (CB). In most cases, the active semiconducting layer is
made of pentacene (black lines) and, in the following, we focus on
these cases.

The pentacene-based transistors differ in the choice of the gate
dielectric and also in the purity of the starting material
(Table~\ref{table-TFTs}). In addition, the trap densities were
calculated with different methods which are also listed in
Table~\ref{table-TFTs}. The specific deviations due to the use of
different methods were discussed in Sec.~\ref{section-quanti}
(Fig.~\ref{figure-compmethods}). These deviations need to be
considered when comparing data obtained with different methods.
Considering these specific deviations, we can draw several
conclusions from Fig.~\ref{figure-TFTs} and Table~\ref{table-TFTs}:
The difference in Fig.~\ref{figure-TFTs} between data no.~1 and data
no.~2 is mainly due to the use of method I by \textit{Kalb et al.}
to obtain data no.~1 and the method by \textit{Gr\"unewald et al.}
to obtain data no.~2. In other words, data no.~1 and data no.~2
correspond to transistors with similar trap densities. A similar
trap DOS is reasonable, because the procedure to fabricate the
transistors was nominally identical in both cases (same deposition
chamber, same gate dielectric, same purity of the starting
material). Data no.~7 in Fig.~\ref{figure-TFTs} implies a rather low
trap density although in that case, too, a SiO$_{2}$ gate dielectric
was used (Table~\ref{table-TFTs}). However, the use of the method by
\textit{Lang et al.} (in particular with an effective accumulation
layer thickness as large as $a=10$\,nm instead of $a=7.5$\,nm)
results in a significant underestimation of the trap DOS close to
the valence band edge and data no.~7 is in fact a sample with a
rather large trap density. Data no.~4 was calculated with the method
by \textit{Fortunato et al.} We consult
Fig.~\ref{figure-compmethods} and conclude that the trap densities
in this sample are indeed very low. Interestingly, the corresponding
field-effect mobilities are as high as
1.2\,cm$^{2}$/Vs.\cite{DeAngelisF2006} Since this transistor was
made with as-received pentacene (Aldrich, 97\%, no additional
purification), this low trap density is most probably not due to a
lower density of chemical impurities in the pentacene film. The low
trap density could be due to the PMMA surface being electrically
passive in the sense that it does not cause charge carrier traps due
to particular chemical groups on its surface when being combined
with pentacene. On the other hand, the growth of pentacene on PMMA
might be exceptionally good thus leading to films with few
structural defects, e.g. at grain boundaries.

\subsection{Trap DOS from SC-FET's}

\begin{figure}
\includegraphics[width=0.90\linewidth]{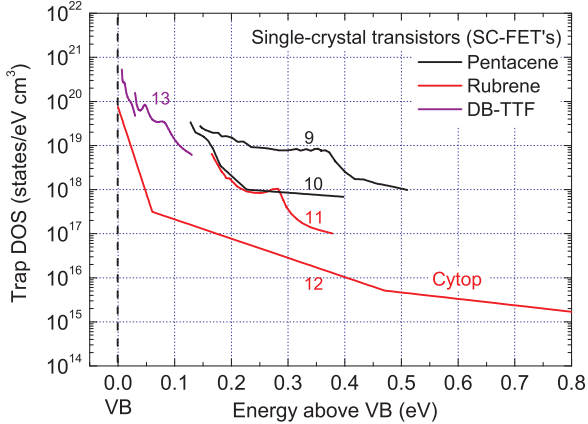}
\caption{\label{figure-SCFETs} (Color online) Trap DOS from single
crystal field-effect transistors (SC-FET's). Different small
molecule semiconductors and gate dielectrics as well as different
calculation methods were used. Details of the data are summarized in
Table~\ref{table-SCFETs}. Remarkable is the very low trap density
from the rubrene-based SC-FET with Cytop$^{TM}$ fluoropolymer gate
dielectric (data no.~12). }
\end{figure}

\begin{table*}
\caption{\label{table-SCFETs} Single crystal field-effect
transistors (SC-FET's) made with small molecule semiconductors:
details of the data shown in Fig.~\ref{figure-SCFETs}. The purity of
the starting material, the gate dielectric, contact effects and the
choice of the method to calculate the trap DOS are expected to
effect the magnitude and slope of the trap distribution. In each
case the original study is cited.}
\begin{ruledtabular}
\begin{tabular}{llllllllll}
Data no. & Semiconductor & Starting material & Gate dielectric &
Contact & Contact  & Method &
Comment & Ref. data\\
&  & & & material & type & &&&  \\  \hline

9 & Pentacene & Aldrich, & Parylene\footnotemark[1]& Colloidal &
BC\footnotemark[2] & \textit{Lang} & $a=7.5$\,nm\footnotemark[3]
& \cite{LangDV2004, ButkoVY2003}\\
&& $2\times$ recrystallized & & graphite or & & &&\\
&&  && silver &  &&&\\

10 & Pentacene & $4\times$ recrystallized &
Parylene\footnotemark[1]& Colloidal & BC & \textit{Lang} &
$a=7.5$\,nm
& \cite{SoWY2008}\\
&&  & & graphite or & & &&\\
&&&& silver &  &&&\\

11 & Rubrene & Aldrich, & Parylene\footnotemark[1]& Colloidal & BC &
\textit{Lang} &
& \cite{SoWY2007}\\
&& recrystallized & & graphite &   & &&\\

12 & Rubrene & & Cytop\footnotemark[4] & Au & BC &
\textit{Oberhoff}\footnotemark[5] & &
\cite{PernstichKPDiss, PernstichKP2008}\\

13 & DB-TTF\footnotemark[6] & & HMDS\footnotemark[7]- or & Pt or Au
& BC & \textit{Lang} & Crystals
grown & \cite{LeufgenM2008}\\
&  &  &
OTS\footnotemark[8]-treated & & & &from solution, & \\
& & & SiO$_{2}$\footnotemark[4] & & &  &$a=7.5$\,nm & \\

\end{tabular}
\end{ruledtabular}

\footnotetext[1]{Top gate transistor structure.} \footnotetext[2]{BC
= bottom contacts (located at the insulator-semiconductor
interface).} \footnotetext[3]{$a$ is the constant effective
accumulation layer thickness used for the
calculations.}\footnotetext[4]{Bottom gate transistor structure.}
\footnotetext[5]{Computer simulations.}
\footnotetext[6]{Dibenzo-tetrathiafulvalene}
\footnotetext[7]{Hexamethyldisilazan.}
\footnotetext[8]{Octadecyltrichlorosilane.}

\end{table*}

In Fig.~\ref{figure-SCFETs} we show the trap DOS in SC-FET's
employing several different small molecule semiconductors. Apart
from data no.~13, all crystals were grown by physical vapour
transport and were either made of pentacene or rubrene (black or red
lines, respectively). The single crystals in these transistors are
grown separately and are then laminated onto the gate dielectric.
This means that the gate dielectric cannot effect the growth of the
organic semiconductor and can thus not be held responsible for
structural disorder within the semiconductor as in the case of
TFT's. For data no.~13 the crystal was grown from solution. Details
of the SC-FET's in Fig.~\ref{figure-SCFETs} are given in
Table~\ref{table-SCFETs}. Data no.~12 stems from simulating the
transistor characteristics with the computer programm developed by
\textit{Oberhoff et al.} and all other trap densities were
calculated with the method by \textit{Lang et al.}

Data no.~9 and no.~10 are from pentacene-based SC-FET's and the gate
dielectric is made of parylene in both cases. The same method was
used to calculate the trap DOS (\textit{Lang et al.}, $a=7.5$\,nm)
and the only difference is that for data no.~9 the starting material
was twice recrystallized and for data no.~10 it was $4\times$
recrystallized. One would conclude, that chemical impurities in the
single crystals have a considerable effect on the magnitude of the
trap densities in SC-FET's at least in the case of
pentacene.\cite{SoWY2008} The method by \textit{Lang et al.} tends
to underestimate the trap DOS, particularly closer to the valence
band edge, i.e. gives lower trap densities as compared to e.g. the
computer simulations with the program developed by \textit{Oberhoff
et al} (see Fig.~\ref{figure-compmethods}). This means that data
no.~12 (obtained with the method by \textit{Oberhoff et al.}) does
indeed correspond to a SC-FET with an extremely low trap density.
This SC-FET employs a rubrene single crystal which was grown in the
usual way (physical vapor transport). However, the transistor
employs a Cytop$^{TM}$ fluoropolymer gate dielectric. Cytop films
are highly water repellent (static water contact angles up to
$116^{\circ}$) and have a very low dielectric constant of
$2.1-2.2$.\cite{KalbWL2007} This strongly suggests that dipolar
disorder due to the presence of the gate dielectric and, more
specifically, water adsorbed on the gate dielectric is a very
important cause of traps in SC-FET's made with an small molecule
organic semiconductor such as rubrene. It appears that water can
cause traps with a wide range of trapping depths.

\subsection{Trap DOS in the bulk of single crystals}

\begin{figure}
\includegraphics[width=0.90\linewidth]{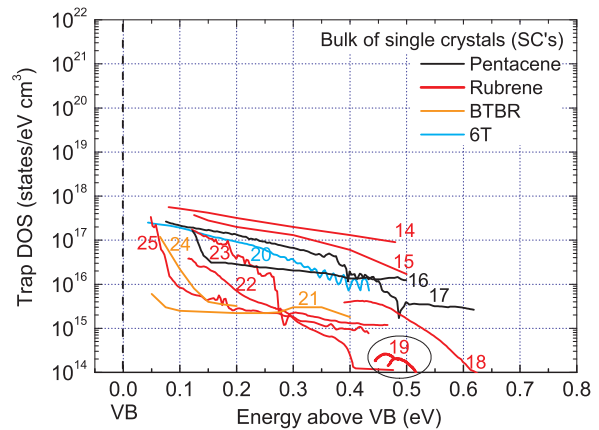}
\caption{\label{figure-bulk} (Color online) Trap DOS in the bulk of
single crystals. The trap densities were calculated from SCLC
measurements and the small molecule semiconductor is e.g. pentacene,
rubrene or sexithiophene. Details of the underlying SCLC
measurements and samples are summarized in Table~\ref{table-SCLC}.}
\end{figure}

\begin{table}
\caption{\label{table-SCLC} Bulk of single crystals made of small
molecule semiconductors: details of the trap DOS data in
Fig.~\ref{figure-bulk}. The trap DOS was calculated from SCLC
measurements in the original studies. Apart from data no.~19,
TD-SCLC measurements (Temperature-dependent SCLC measurements,
Refs.~\onlinecite{SchauerF1996}, \onlinecite{SchauerF1997} and
\onlinecite{KrellnerC2007}) and a sandwich-type device structure
were employed. In all cases Au electrodes were used.}
\begin{ruledtabular}
\begin{tabular}{llll}
Data no. & Semiconductor & Starting material & Ref. data\\
\hline

14 & Rubrene & Aldrich (purum), & \cite{KrellnerC2007}\\
& & $7\times$ recrystallized & \\

15 & Rubrene & Aldrich (purum), & \cite{KrellnerC2007}\\
& & $3\times$ recrystallized & \\

16 & Pentacene & &\\

17 & Pentacene & $4\times$ recrystallized & \\

18 & Rubrene & &\\

19\footnotemark[1]\footnotemark[2] & Rubrene & & \cite{BragaD2008}\\

20 & 6T\footnotemark[3] & $2\times$ recrystallized & \\

21 & 5,11-BTBR (B)\footnotemark[4] &  Ciba SC & \cite{HaasS20072}\\

22 & Rubrene & & \\

23 & Rubrene & Aldrich (purum), & \cite{KrellnerC2007}\\
& & $3\times$ recrystallized & \\

24 & 5,11-BTBR (B) &  Ciba SC & \cite{HaasS20072}\\

25 & Rubrene & Aldrich (purum), & \cite{KrellnerC2007}\\
& & $3\times$ recrystallized & \\

\end{tabular}
\end{ruledtabular}
\footnotetext[1]{DM-SCLC (Differential-method SCLC,
Ref.~\onlinecite{NespurekS1980} and
\onlinecite{BragaD2008}).}\footnotetext[2]{Coplanar (gap)
structure.} \footnotetext[3]{Sexithiophene.}
\footnotetext[4]{Rubrene derivative t-butyl-tetraphenylrubrene,
polymorph B.}
\end{table}

In Fig.~\ref{figure-bulk} we show the trap DOS in the bulk of single
crystals made of small molecule semiconductors. The hole trap
densities were calculated from SCLC measurements. In all cases, the
crystals were grown by physical vapour transport. Some details of
the data are given in Table~\ref{table-SCLC}. Apart from one
exception, all data were obtained with samples that had a
sandwich-type structure (contact-crystal-contact) and
temperature-dependent SCLC measurements (TD-SCLC) were
used.\cite{SchauerF1996, SchauerF1997, KrellnerC2007} For data
no.~19 however, a coplanar gap structure (both contacts on the same
side of the crystal) and differential method SCLC (DM-SCLC) was
employed.\cite{NespurekS1980, BragaD2008} Au electrodes were used in
all cases. In many cases, rubrene crystals were measured (red lines
in Fig.~\ref{figure-bulk}) and in the following we focus on the trap
DOS in the bulk of rubrene single crystals.

Data no.~15, 23 and 25 are the trap DOS in three different rubrene
crystals but for all of these crystals, the starting material was
$3\times$ recrystallized. Since we have significant differences in
the trap densities when comparing these crystals but the same purity
of the starting material, we conclude that it is not chemical
impurities but structural defects that are the main cause of traps
in the bulk of rubrene crystals grown by physical vapour transport.
This would be as in the case of ultrapure (zone-refined) crystals
e.g. made of naphtalene.\cite{WartaW1985} At least, we cannot
identify any clear correlation between the magnitude of the trap
densities and the number of the recrystallization steps to purify
the starting material. Interestingly, the highest trap densities
were obtained when the rubrene crystals were grown with starting
material that had been recrystallized most often ($7\times$
recrystallized, data no.~14). This further supports the dominance of
structural defects in the bulk of rubrene crystals.

\subsection{Comparison: TFT's, SC-FET's and bulk}

\begin{figure*}
\includegraphics[width=0.7\linewidth]{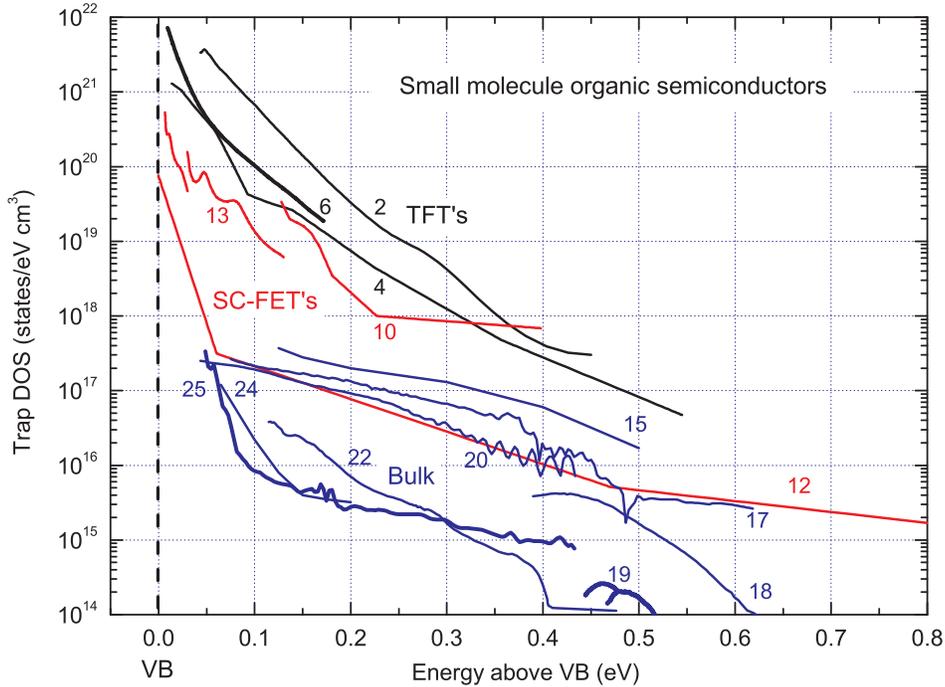}
\caption{\label{figure-together} (Color online) Representative trap
DOS data in small molecule organic semiconductors from thin-film
transistors (TFT's, black lines), single-crystal field-effect
transistors (SC-FET's, red lines) and in the bulk of single crystals
(blue lines). The data were selected from Fig.~\ref{figure-TFTs},
\ref{figure-SCFETs} and \ref{figure-bulk} as typical examples. The
trap densities from SC-FET's can be much lower than the trap
densities from TFT's. This strongly suggests that traps due to
structural defects tend to dominate in thin films. The trap
densities in the bulk of single crystals are typically lower than
the trap densities from SC-FET's. Importantly, if a Cytop$^{TM}$
fluoropolymer gate dielectric is used, bulk trap densities can be
reached in organic field-effect transistors made with small molecule
semiconductors. Thus, water adsorbed on the gate dielectric appears
to be the dominant cause of traps if the semiconductor has a low
density of traps due to structural defects (e.g. single crystals). A
steep increase of the trap DOS very close to the valence band edge
($<0.15$\,eV) can sometimes be observed (data no. 6, 12, 24 and 25).
These states are attributed to the thermal fluctuations of the
intermolecular transfer integral.}
\end{figure*}

In Fig.~\ref{figure-together} we show typical trap densities in
TFT's (black lines), SC-FET's (red lines) and in the bulk of single
crystals (blue lines). The data were selected from
Fig.~\ref{figure-TFTs}, \ref{figure-SCFETs} and \ref{figure-bulk} as
typical examples.

The trap densities in SC-FET's and in the bulk of single crystals
can be much lower than the trap densities in TFT's. We conclude that
growth-related structural defects tend to be the main cause of traps
in TFT's made with small molecule semiconductors such as pentacene.
These structural defects are likely to be concentrated at grain
boundaries: According to Ref.~\onlinecite{VerlaakS2007}, in-grain
structural defects cannot exceed $10^{16}-10^{17}$\,cm$^{-3}$ at
typical growth conditions. In all cases, the trap DOS is calculated
from current-voltage characteristics by assuming that we have an
infinite electrical stability and, e.g., no current hysteresis. This
means that the present study deals with ``fast'' traps, i.e. traps
with trapping and release times much shorter than e.g. the time to
measure a transistor characteristic (e.g. 1\,min.). Interestingly,
in the case of pentacene-based thin-film transistors, long-lived and
energetically deep traps ($>0.5$\,eV from the valence band edge)
that cause gate bias stress effects are mainly located at grain
boundaries as well.\cite{TelloM2008}

When comparing the trap densities in SC-FET's with the trap
densities in the bulk of single crystals, we see that the trap
densities are typically lower in the bulk. However, for the
rubrene-based SC-FET with the highly hydrophobic Cytop fluoropolymer
gate dielectric (data no.~12 in Fig.~\ref{figure-together}) the trap
densities are comparable to the trap densities in the bulk of some
rubrene crystals. Consequently, bulk trap densities can be reached
in organic field-effect transistors if the organic semiconductor has
few structural defects (e.g. single crystals, no grain boundaries)
and if a highly hydrophobic gate dielectric is used. Water adsorbed
on the gate dielectric appears to be the main cause of traps in
SC-FET's and can cause traps in a wide range of energies.

If we only consider the trap densities in Fig.~\ref{figure-together}
for energies $>0.15$\,eV, the magnitude of the trap densities
appears to be correlated with the steepness of the trap
distribution. The steepest trap distributions are present in TFT's.
For example, fitting data no.~2 to an exponential function
$N=N_{0}\exp(-E/E_{0})$ yields $E_{0}=32$\,meV.\cite{KalbWL20072}
The trap DOS is significantly less steep in the bulk of organic
crystals (e.g. data no.~25: $E_{0}=180$\,meV,
Ref.~\onlinecite{KrellnerC2007}). The explanation for this
correlation is not clear at present.

Interestingly, in several samples a steep increase of the trap DOS
very close to the valence band edge (for energies $<0.15$\,eV) can
be observed, especially in samples with a low trap density (data no.
6, 12, 24 and 25). These traps are of particular importance for the
performance of organic field-effect transistors. We offer two
different explanations for the steep increase in the trap DOS close
to the valence band edge. On the one hand, these traps may be the
signature of thermal fluctuations of the intermolecular transfer
integral.\cite{SleighJP2009} The thermal motion of the small
molecules are expected to result in an exponential tail of trap
states and calculations predict $E_{0}=10-20$\,meV at
$T=300$\,K.\cite{SleighJP2009} From experiment we have
$E_{0}=22$\,meV (data no.~24, Ref.~\onlinecite{HaasS20072}), 11\,meV
(data no.~25, Ref.~\onlinecite{KrellnerC2007}) and 11\,meV (data
no.~12, Ref.~\onlinecite{PernstichKP2008}). Although the agreement
between theory and experiment is compelling, we keep in mind that
contact effects can be significant in organic semiconductor devices.
Good electrical contacts to an organic semiconductor are difficult
to be achieved.\cite{AnthonyJE2008} For example, contact resistances
at the source and drain contact of an organic field-effect
transistor are often neglected when calculating the trap DOS but can
lead to an overestimation of the trap DOS particularly very close to
the valence band edge.\cite{KalbWL2008} With the existing data we
cannot completely rule out the possibility that the steep increase
in the trap DOS is an artifact of non-ideal (limiting) contacts.

\subsection{Oxygen-related traps}

\begin{figure}
\includegraphics[width=0.90\linewidth]{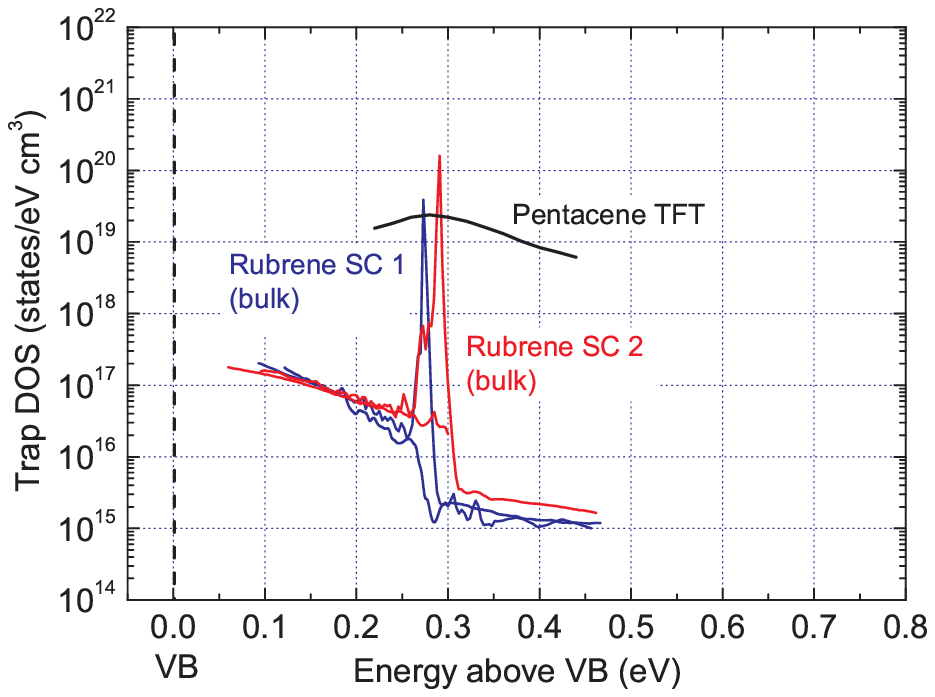}
\caption{\label{figure-oxygen} Oxygen-induced traps in rubrene and
pentacene. The exposure of rubrene single crystals (SC) to oxygen in
combination with light leads to a sharp peak of trap states centered
at about 0.28\,eV (two different samples, blue and red lines, data
from Ref.~\onlinecite{KrellnerC2007}). Exposing pentacene thin films
(TFT's) to oxygen in combination with light leads to a much broader
peak of trap states also centered at 0.28\,eV (full black line, data
from Ref.~\onlinecite{KalbWL2008}). For the pentacene thin film, the
width of the peak is 0.16\,eV and the volume density of traps as
calculated from integrating the peak is $4\times10^{18}$\,cm$^{-3}$.
The large width of the peak is thought to result from the increased
local disorder in a thin film as compared to the bulk of single
crystals.}
\end{figure}

We now discuss the effect of oxygen-related chemical impurities on
the trap DOS. In Fig.~\ref{figure-oxygen} we compare the effect of
oxygen exposure (in combination with light) on the trap DOS of two
rubrene crystals (red and blue lines) and on the trap DOS of a
pentacene thin film (black line).\cite{KrellnerC2007, KalbWL2008}
For the rubrene crystals, TD-SCLC measurements were used. The peak
in the trap DOS of the pentacene film was determined by employing
TFT measurements and method I by \textit{Kalb et al.}

Oxygen in combination with light results in oxygen radicals that
react with the organic semiconductors. For two different organic
semiconductors (pentacene and rubrene) the oxygen exposure results
in a peak that is centered at the same energy, i.e. at about
0.28\,eV.

For rubrene crystals, oxygen exposure leads to a sharp peak in the
trap DOS. In the case of pentacene films, we have a peak with a very
large width of 0.16\,eV with a total concentration of states of
order $10^{18}$\,cm$^{-3}$.\cite{KalbWL2008} The large width of the
peak is thought to result from the increased local structural
disorder in a thin film. The disorder modifies the on-site energy of
the oxygen-affected molecules and leads to a broadening of the
peak.\cite{KalbWL2008}

Theoretical studies predict various types of oxygen-related defects
in pentacene.\cite{NorthrupJE2003, TsetserisL2007, BenorA2008,
KnippD2009} In Ref.~\onlinecite{NorthrupJE2003} oxygen defects are
discussed in which a H atom of a pentacene molecule is replaced by
an oxygen atom to form a C$_{22}$H$_{13}$O molecule. The oxidation
at the middle ring (6- or 13-position) of the pentacene molecule is
shown to be energetically most favorable.\cite{NorthrupJE2003} The
oxidation of the middle ring at one of the two sites results in the
formation of two trap states in the bandgap of pentacene. These are
located at 0.18\,eV and 0.62\,eV above the valence band
maximum.\cite{BenorA2008} In Ref.~\onlinecite{TsetserisL2007} other
oxygen defects in pentacene are described. An example is a single
oxygen intermolecular bridge where a single oxygen atom is
covalently bound to the carbon atoms on the center rings of two
neighboring pentacene molecules. This defect, for instance, is
calculated to lead to electrically active traps at 0.33 and 0.4\,eV
above the valence band maximum.\cite{TsetserisL2007} In
Ref.~\cite{KnippD2009} similar defects are described: An O$_{2}$
molecule may dissociate and the two oxygen atoms are bound at the 6-
and 13-positions of the pentacene molecule. Calculations predict a
very shallow state at 0.08\,eV above the valence band
maximum.\cite{KnippD2009} However, this pentacene complex can reduce
its energy if one of the oxygen atoms forms a bond with a carbon
atom of a neighboring pentacene molecule. This leads to
acceptor-like states (0/-) at 0.29\,eV above the valence band
maximum.\cite{KnippD2009} The experimentally observed effect of
oxygen-exposure in combination with gate bias stress at positive
gate voltages on the transfer characteristics of pentacene TFT's can
be modeled by introducing a Gaussian distribution of acceptor-like
states at 0.29\,eV with a width of 0.1\,eV and a total concentration
of the order of $10^{18}$\,cm$^{-3}$.\cite{KnippD2009}


\subsection{Comparison with hydrogenated amorphous and polycrystalline silicon}

\begin{figure}
\includegraphics[width=0.90\linewidth]{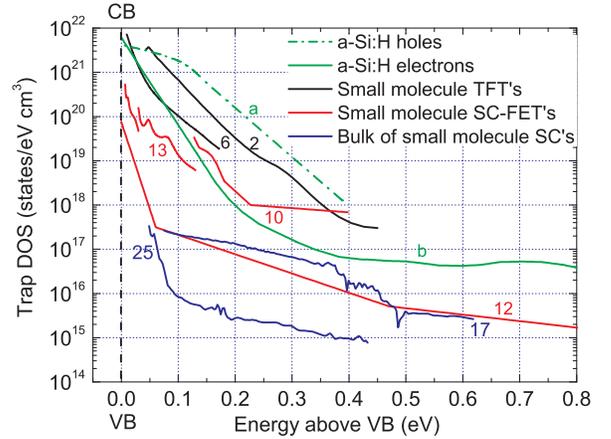}
\caption{\label{figure-aSi} Comparison of typical trap densities in
small molecule organic semiconductors/transistors with typical trap
densities in hydrogenated amorphous silicon (a-Si:H). The
dashed-dotted green line is a typical distribution of hole traps in
a-Si:H (energy relative to VB). The full green line marks typical
electron trap densities in a-Si:H (energy relative to CB). Details
of the data are given in Table~\ref{table-Si}. The hole trap DOS in
a-Si:H is surprisingly similar to the hole trap DOS in
small-molecule-based TFT's.}
\end{figure}

\begin{figure}
\includegraphics[width=0.90\linewidth]{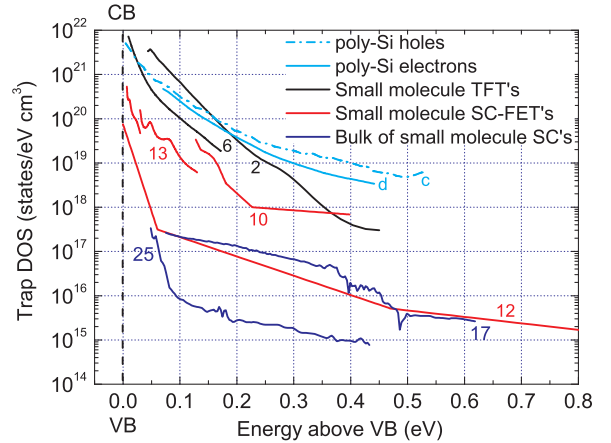}
\caption{\label{figure-polySi} Comparison of typical trap densities
in small molecule semiconductors with typical trap densities in
polycrystalline silicon (poly-Si). The dash-dotted blue line
represents hole traps in poly-Si (energy relative to VB) and the
full blue line marks electron traps in poly-Si (energy relative to
CB). Details of the data are given in Table~\ref{table-Si}.}
\end{figure}

\begin{table}
\caption{\label{table-Si} Details of the trap DOS data in
Fig.~\ref{figure-aSi} from hydrogenated amorhous silicon (a-Si:H)
and in Fig.~\ref{figure-polySi} from polycrystalline silicon
(poly-Si) samples.}
\begin{ruledtabular}
\begin{tabular}{lllll}
Data & Semiconductor & Carriers & Based on & Ref. data\\
\hline

a & a-Si:H,& Holes & Photoemission &
\cite{StreetRA1991}, \\
& good quality& & and time-of-flight & (p.~81)\\


b & a-Si:H & Electrons & TFT & \cite{DjamdjiF19872}\\

c & poly-Si & Holes & TFT & \cite{ChernHN1994}\\

d & poly-Si & Electrons & TFT\footnotemark[1] & \cite{FortunatoG1988}\\

\end{tabular}
\end{ruledtabular}
\footnotetext[1]{Calculation method: \textit{Fortunato et al.}}
\end{table}

It is interesting to compare the trap DOS in small molecule organic
semiconductors with the trap DOS in hydrogenated amorphous silicon
(a-Si:H) and polycrystalline silicon (poly-Si). For a-Si:H, the
mobility edge picture is used to describe the charge transport and
trap states have been studied extensively.\cite{MarshallJM1983,
StreetRA1991} The distribution of bond angles and interatomic
distances in amorphous silicon (a-Si) around a mean value leads to a
blurred band edge, i.e. to band tails extending into the gap. The
trap densities at a given energy reflect the volume density of
certain bond angles and interatomic distances. For example, a rather
large deviation from the atomic configuration in the crystalline
phase (from the mean value in the amorphous phase) leads to traps
with energies far from the band edge. These traps are present with
rather low densities since small deviations are much more likely to
occur. In addition, we may have dangling bonds in a-Si acting as
traps. It is well known, that hydrogenation of a-Si leads to a
reduction in the trap DOS due to a passivation of dangling bonds
with hydrogen.\cite{StreetRA1991}

For Fig.~\ref{figure-aSi} we have selected typical trap DOS data
from samples with small molecule semiconductors (data from
Fig.~\ref{figure-together}). The data are compared with a typical
hole trap DOS in a-Si:H (dash-dotted green lines) and with a typical
electron trap DOS in a-Si:H (full green line). Details of the data
are given in Table~\ref{table-Si}. In Fig.~\ref{figure-aSi} we see
that the hole trap DOS in TFT's with small molecule semiconductors
such as pentacene is surprisingly similar to the hole trap DOS in
a-Si:H. Both the magnitude of the trap densities and the slope of
the distribution are very similar.

Finally, in Fig.~\ref{figure-polySi} we similarly compare data from
small molecule semiconductors with a typical hole trap DOS in
poly-Si (dash-dotted blue line) and an electron trap DOS in poly-Si
(full blue line). The trap distribution is less steep in poly-Si as
compared to the trap DOS in organic thin films such that we have
higher trap densities far from the transport band edge.

\section{Summary and conclusions}

We compared the hole trap DOS (trap densities as a function of
energy relative to the valence band edge) in various samples of
small molecule organic semiconductors as derived from electrical
characteristics of organic field-effect transistors and
space-charge-limited current measurements. In particular, we
distinguish between the trap DOS in thin-film transistors with
vacuum-evaporated small molecules, the trap DOS in organic single
crystal field-effect transistors and the trap DOS in the bulk of
single crystals grown by physical vapour transport. A comparison of
all data strongly suggests that structural defects at grain
boundaries tend to be the main cause of ``fast'' traps in TFT's made
with vacuum-evaporated pentacene and supposedly also in related
materials. Moreover, we argue that dipolar disorder due to the
presence of the gate dielectric and, more specifically, water
adsorbed on the gate dielectric surface is the main cause of traps
in SC-FET's made with a semiconductor such as rubrene. One of the
most important findings is that bulk trap densities can be reached
in organic field-effect transistors if the organic semiconductor has
few structural defects (e.g. single crystals) and if a highly
hydrophobic gate dielectric is used. The highly hydrophobic
Cytop$^{TM}$ fluoropolymer gate dielectric essentially is a gate
dielectric that does not cause traps at the insulator-semiconductor
interface and thus leads to organic field-effect transistors with
outstanding performance.

The trap DOS in TFT's with small molecule semiconductors is very
similar to the trap DOS in hydrogenated amorphous silicon. This is
surprising due to the very different nature of polycrystalline thin
films made of small molecule semiconductors with van der Waals-type
interaction on the one hand and covalently bound amorphous silicon
on the other hand.

Although several important conclusions can be drawn from the
extensive data it is clear that the present picture is not complete.
More systematic studies are necessary to consolidate and complete
the understanding of the trap DOS in organic semiconductors and
organic semiconductor devices. The present compilation may serve as
a guide for future studies.

\begin{acknowledgments}
The authors thank Kurt Pernstich for a careful reading of the
manuscript and for valuable suggestions. We thank Kurt Mattenberger
for constant technical support.
\end{acknowledgments}



\end{document}